\documentclass[sigconf,nonacm,balance=false]{acmart}

\makeatletter
\newif\ifanonymous
\@ifclasswith{acmart}{anonymous}{\anonymoustrue}{\anonymousfalse}
\makeatother

\usepackage[british]{babel}
\usepackage[utf8]{inputenc}

\def\plaintitle{GDPR: Is it worth it? Perceptions of workers who have experienced its implementation}

\ifanonymous \def\plainauthor{Anonymous for submission} \else \def\plainauthor{Gerard Buckley, Tristan Caulfield, Ingolf Becker} \fi

\def\plainkeywords{GDPR; Data protection; Regulation; Perceptions; Trade-offs}
\date{Feb 2024}

\usepackage{hyperref}
\hypersetup{%
 pdftitle={\plaintitle},
 pdfauthor={\plainauthor},
 pdfkeywords={\plainkeywords},
 pdfdisplaydoctitle=true, %
 bookmarksnumbered,
 pdfstartview={FitH},
 breaklinks=true,
 colorlinks=true,
 allcolors=black,
 hypertexnames=false
}
\usepackage{pdfpages}

\usepackage{booktabs}
\usepackage[export]{adjustbox}
\usepackage{multirow}
\usepackage{graphicx}
\graphicspath{{img/}}

\usepackage{booktabs} %
\usepackage{lipsum}
\usepackage{balance}
\usepackage{diagbox}

\usepackage{enumitem}
\setlist{noitemsep}

\usepackage{tikz}
\usetikzlibrary{backgrounds}
\usetikzlibrary{arrows}
\usetikzlibrary{shapes,shapes.geometric,shapes.misc}

\usepackage{xcolor}
\usepackage[style=british]{csquotes}

\DeclareQuoteStyle[british]{english}
  {\textquoteleft}%
  {\textquoteright}
  [0.05em]
  {\textquotedblleft}
  {\textquotedblright}

\DeclareQuoteStyle[american]{english}
  {\itshape\textquotedblleft}
  {\textquotedblright}
  [0.05em]
  {\textquoteleft}
  {\textquoteright}

\SetBlockEnvironment{quotation} 

\renewcommand{\mkbegdispquote}[2]{\openautoquote}

\newcolumntype{C}[1]{>{\centering\arraybackslash}m{#1}}

\newlength{\sLQ}
\newlength{\sLP}
\newlength{\sLPsep}
\newlength{\lmQ}
\newlength{\lmP}
\newlength{\lmPsep}
\newlength{\lmIQ}
\newlength{\lmIP}
\newlength{\lmIPsep}
\begin{document}

\title[GDPR: Is it worth it?]{GDPR: Is it worth it?\\ Perceptions of workers who have experienced its implementation}

\author{Gerard Buckley}
\affiliation{%
  \institution{University College London}\country{UK}
}
\email{gerard.buckley.18@ucl.ac.uk}

\author{Tristan Caulfield}
\affiliation{%
  \institution{University College London}\country{UK}
}
\email{t.caulfield@ucl.ac.uk}

\author{Ingolf Becker}
\affiliation{%
  \institution{University College London}\country{UK}
}
\email{i.becker@ucl.ac.uk}

\renewcommand{\shortauthors}{Buckley, Caulfield \& Becker}



\begin{abstract}
The General Data Protection Regulation (GDPR) remains the gold standard in privacy and security regulation. We investigate how the cost and effort required to implement GDPR is viewed by workers who have also experienced the regulations' benefits as citizens: is it worth it? In a multi-stage study, we survey N = 273 \& 102 individuals who remained working in the same companies before, during, and after the implementation of GDPR. The survey finds that participants recognise their rights when prompted but know little about their regulator. They have observed concrete changes to data practices in their workplaces and appreciate the trade-offs. They take comfort that their personal data is handled as carefully as their employers' client data. The very people who comply with and execute the GDPR consider it to be positive for their company, positive for privacy and not a pointless, bureaucratic regulation. This is rare as it contradicts the conventional negative narrative about regulation. 
Policymakers may wish to build upon this public support while it lasts and consider early feedback from a similar dual professional-consumer group as the GDPR evolves.
\end{abstract}


\maketitle

\section{Introduction}

People who were employed before May 2018 and who are still employed by the same organisation will have experienced the impact of the GDPR on their workplace firsthand. They both implement it as employees and benefit from it as consumers. The goal of this study is to understand the unique dual perspective of this group.

The GDPR has been studied from multiple points of view. It ranges from the implementation challenges business face~\cite{poritskiy_benefits_2019a, buckley_it_2022} to the enforcement issues Data Protection Authorities (DPA) face~\cite{johnnyryanalantoner_new_2020,grant_how_2016, johnson_economic_2022, chazal_20_} to the operational realities that consumers face~\cite{nouwens_dark_2020}. The European Commission (EC) and professional services firms have surveyed consumers' awareness of their rights and businesses' awareness of their obligations. In academia, there have been reactance studies~\cite{strycharz_data_2020b} and comparative awareness studies across Europe~\cite{rughinis_social_2021a}. Unlike previous perception studies that focused solely on consumers or data professionals, this is the first empirical research into how these informed individuals perceive the cost-benefit of their rights as consumers balanced against the pressures they see it places on their employer to support those rights. Hence our research question---GDPR: Is it worth it?

To exercise their rights, consumers need to be aware, to some extent, of the regulator's identity, role, and powers. The EC~\cite{_gdpr_2019,europeancommission.directorategeneralforjusticeandconsumers._general_2019c,belgiandpa_belgian_2021,greatbritain_information_2021a} and some DPAs have conducted consumer awareness and confidence surveys but we find little evidence of systematic overt publicity campaigns. We test if our informed citizens know who their regulator is and what they expect of them.

Most business-focused coverage of the GDPR in the media concentrates on data breaches and regulators' fines~\cite{chazal_20_, wolff_early_2020, venkataramakrishnan_gdpr_2021}. It stresses the deterrence effects of the GDPR sanctions at the expense of any incentive to change or upside for business. If the point of privacy regulation is behaviour change~\cite{carycoglianese_measuring_2012, theenvironmentagency_effectiveness_2011}, measuring it is difficult. Available data, such as the number of fines, delivers a highly imperfect and incomplete picture of compliance within companies. Instead, we test what GDPR-driven changes have been observed by our respondents within their organisation and if they believe these changes have been net-positive.

At the end of the survey, after making our respondents consider the GDPR from multiple angles, we ask if they feel the GDPR has been worth it. Their answer is important because it cuts to the heart of privacy in the digital age. Without data protection, citizens will arguably be exposed to more profiling, monitoring and mass influencing by digital advertisers and/or the state. We find the informed citizen-consumer does buy into the GDPR, with all its positives and negatives. This has important implications for policymakers and regulators who may wish to learn how to emulate this public support and build on it for future regulation roll-outs.

\section{Background to the GDPR}
This section does not aim to give a detailed description of the GDPR. Rather, we focus on the  big themes and provide summaries of provisions that are relevant to our area of interest. For a more fundamental introduction, please see Voigt \& Bussche~\citeyear{voigt_eu_2017} or Hoofnagle et al.~\citeyear{hoofnagle_european_2019b}. Bear in mind that despite Brexit, the UK GDPR is essentially equivalent to the EU GDPR~\cite{europeancommission_commission_2021}.

Europe has long recognised privacy as a fundamental or human right. Article 8 of the European Convention on Human Rights provides a right to respect for one's `private and family life, his home and his correspondence', subject to certain restrictions that are `in accordance with law' and `necessary in a democratic society'~\cite{echr_guide_2021}.
In contrast, the Constitution of the United States and the United States Bill of Rights do not explicitly contain a right to privacy. Instead, there is an implied right to privacy derived from penumbras of other explicitly stated constitutional protections~\cite{cornelllawschool_privacy_,ussupremecourt_estelle_1965}.

While much of the focus of US law is the home, EU law has taken a wider approach. Where U.S. law may refer broadly to `privacy’ or to `information privacy’, EU law discusses information privacy as `data protection'. In Europe, data protection and the right to privacy are beginning to be viewed as separate. Data protection focuses on the usage, storage and movement of data while privacy preserves the Athenian concept of private and public life~\cite{hoofnagle_european_2019b}. 

Under the GDPR, companies that use personal data have to follow strict rules called `data protection principles'. They must make sure the information is: used fairly, lawfully and transparently;
used for specified, explicit purposes;
used in a way that is adequate, relevant and limited to only what is necessary;
accurate and, where necessary, kept up to date;
kept for no longer than is necessary;
handled in a way that ensures appropriate security, including protection against unlawful or unauthorised processing, access, loss, destruction or damage;
There is stronger legal protection for more sensitive information, such as race, religion, etc~\cite{ico_data_2018}.

Under the GDPR, individuals have the right to find out what information the government and other organisations store about them. These include the right to: be informed about how their data is being used;
access personal data;
have incorrect data updated;
have data erased;
stop or restrict the processing of their data;
data portability (allowing individuals to get and reuse their data for different services);
object to how their data is processed in certain circumstances;
Individuals also have rights when an organisation is using their personal data for: automated decision-making processes (without human involvement);
profiling, for example to predict behaviour or interests~\cite{ico_data_2018}.

The direction of these rules is clear: companies will have to take more responsibility for the information they handle and hold, and individuals will be empowered to gain more control over their own data. The following literature review will survey prior works on these two topics and their interplay. 

\section{Literature review}

Given the unique nature of our research target group, we review the literature from a consumer and business perspective---namely consumer awareness and knowledge of the GDPR and DPAs, consumer perceptions of privacy, the response of business to implementing the GDPR, the awareness of staff within the business of the measures required to operationalise it and their perception of its benefit to them and to their company. We find outstanding contradictions in prior works and blind spots that lead us to a series of research questions that we investigate further in our study as summarised in Section~\ref{sec:summary}.

\subsection{Consumer awareness and knowledge of the regulation}

An informed citizenry is vital for a well-functioning democracy. %
The GDPR makes the awareness-raising duties of Data Protection Authorities (DPA) explicit. Under Article 57.1, the DPAs have an obligation to \enquote{promote public awareness and understanding of the risks, rules, safeguards and rights in relation to processing}. DPAs employ various activities to raise awareness, ranging from publishing press releases, penalty notices, educational materials and commentaries, to hosting public meetings and events. These activities are aimed at companies as much as at citizens~\cite{jasmontaite-zaniewicz_gdpr_2021}.

The success of this awareness-raising challenge attracts wildly differing verdicts.
On the low side, a private survey of over 289K consumers, coinciding with the first anniversary of the GDPR on May 2019, found \enquote{A staggering eight percent of consumers globally feel they have a better understanding of how companies use their data since the GDPR's introduction}~\cite{_gdpr_2019}. %

On the high side, a Eurobarometer survey on the same May 2019 anniversary found 67\% of respondents had heard of the GDPR, 36\% had heard of it and knew what it was, almost 73\% had heard of at least one right guaranteed by the GDPR and 31\% had heard of all the rights asked about in the survey~\cite{europeancommission.directorategeneralforjusticeandconsumers._general_2019c}. The level of awareness varied wildly between countries, from 90\% in Sweden all the way to 44\% in France. 

A later secondary analysis of the same 2019 EU Eurobarometer survey showed education, occupation, and age were the strongest socio-demographic predictors of the GDPR awareness, with little influence of gender, subjective economic well-being, or locality size. 

Sources of information also differ. In the 2020 \enquote{Data Protection or Data Frustration? Individual Perceptions and Attitudes towards the GDPR}, Strycharz et al. found most respondents learnt about the Regulation from the news (47\%), followed by their employer (36\%) and cookie notices on websites~\citeyear{strycharz_data_2020b}.

Despite the contradictory results from prior surveys, we hypothesise the trend is positive.

\noindent \textbf{Hypothesis 1}: Consumers are aware and knowledgeable about the GDPR.

\subsection{Consumer awareness and knowledge of the regulator}

Conscious of their duty to promote public awareness, regulators conduct surveys, albeit their definitions and metrics vary wildly across the EU. Professional services firms and data rights groups also conduct and publish GDPR-related surveys. Academic surveys on regulator awareness are sparse.

An EU Eurobarometer survey~\cite{europeancommission.directorategeneralforjusticeandconsumers._general_2019c} of 27,524 people across the 28 member states found 57\% had heard about the existence of a public authority in their country responsible for protecting their rights regarding their personal data---an increase of 20\% on a 2015 survey.

The Belgian DPA~\citeyear{belgiandpa_belgian_2021} takes a different approach. In its 2020 annual report, it congratulates the increased awareness of citizens because \enquote{the year 2020 saw a sharp increase in the number of complaints ($+ 290.64$\%) and data breach notifications ($+ 25.09$\%) received by the BE DPA and, more generally, a significant increase in the DPA's workload}. 

The UK Information Commissioner's (ICO) 2021 Annual Report ~\citeyear{greatbritain_information_2021a} contains an annual track survey of 2,000 people to measure its strategic performance in supporting the public. It found 28\% of people have high trust and confidence (compared with 27\% in 2020), with a similar number state they have low trust and confidence (29\%, compared with 28\% in 2020). 

Awareness of the GDPR translates for some into awareness of the regulator's punitive power. For example, the international law firm DLA Piper~\citeyear{dlapiper_dla_2022} publishes an annual fine and data breach survey as part of their public relations strategy to communicate the need for proper advice on GDPR matters. Privacy advocates follow a similar path. AccessNow~\cite{estellemasse_threeyearsundergdprreport_2021a} is a digital rights group that publishes an annual evaluation of GDPR. In its most recent report `Three Years Under GDPR', it focuses on the number and value of data fines and concludes `GDPR implementation is proving to be nothing but hot air' due to a lack of enforcement.

In sum, apart from the EC's four-yearly survey, regulators are not measuring their brand awareness. Instead they measure proxies such as customer confidence, consumer complaints or breach notifications whilst professional services firms and advocacy groups focus more on fines for their own reasons. %
Familiarity with the regulator among laypeople appears to be under or untested. We  conjecture:

\noindent \textbf{Hypothesis 2}: Consumers lack awareness and knowledge of the regulator.

\subsection{Consumer perceptions of privacy}

Side-stepping the privacy-paradox debate and whether privacy desires and privacy actions are consistent, we are interested in how people feel or perceive the state of their privacy post-GDPR. We find contradictory studies that claim it has had little impact, and others believe it has improved people's feeling of privacy.

In `Are consumers concerned about privacy?', conducted in the run-up to GDPR, Presthus and Sørum found the respondents had a favourable view of the GDPR, but they were sceptical about its enforcement~\citeyear{presthus_consumer_2019}. In a follow-up, `A three-year study of the GDPR and the consumer', they found that the GDPR has not significantly affected consumer awareness nor the level of control over their own personal data~\citeyear{presthus_threeyear_2021a}.

There is some evidence that consumer perceptions of power and risk in digital information privacy have risen due to mandatory the GDPR cookie notices%
~\cite{bornschein_effect_2020}.  %
In a similar vein Zhang et al.~\citeyear{zhang_online_2020b} suggest that the GDPR plays a significant role in online customer trust by bringing about stronger rights and more transparency for online customers.

A 2021 survey by the ICO~\cite{worledge_information_2021} found that 77\% of people say protecting their personal information is essential. The survey does not ask about privacy per se. Instead, it asked `Has your trust and confidence in companies and organisations storing and using your personal information increased, decreased or stayed the same in the past year?', they found 9\% felt it had increased, 68\% felt it had stayed the same and 23\% felt it had decreased compared to 2020. The answers in percentages were broadly the same as the 2020 survey results. Given the unique perspective of our sample, we believe that it is apposite to seek their evaluation of the effect of the GDPR on their privacy perceptions:

\noindent \textbf{Hypothesis 3}: Consumers feel their privacy is better since the GDPR was introduced.

\subsection{Business response to Data Protection regulation}

To comply with the GDPR, most companies will have had to make some legal, technical and organisational changes~\cite{poritskiy_benefits_2019a}. Failure to comply can attract hefty fines of up to 4\% of global turnover ~\cite{gdpr.eu_what_2018} and negative publicity which in turn can affect a company's market valuation if it is publicly quoted. 

A systematic literature review~\cite{spanos_impact_2016} into the economic consequences of security incidents found that most studies (76\%) report a statistically significant negative impact of data breach events on the stock market. Ford et al.~\citeyear{ford_impact_2021} found cumulative abnormal returns of around $-1\%$ after three days far outweighed the monetary value of the fine itself, and relatively minor fines could result in major market valuation losses for companies. The persistence of this effect is open to debate and Richardson, Smith and Watson argue that `companies are unlikely to change their investment patterns unless the cost of breaches increases dramatically or regulatory bodies enforce change'~\citeyear{richardson_much_2019}.

Buckley et al.~\citeyear{buckley_it_2022} found the fear of the GDPR's threat of meaningful financial penalties has spurred companies to take the GDPR seriously. It has led to modernisation of company databases, more careful accounting of data, and greater awareness of information security. %
Cochrane et al. and Jasmontaite-Zaniewicz et al.~\citeyear{jasmontaite-zaniewicz_gdpr_2021,cochrane_data_2020} surveyed SME Associations and found evidence to support Hijmans'~\citeyear{hijmans_how_2018} view that information and awareness of the imposition of fines was a regularly cited way of capturing the attention of SMEs.

Thus GDPR compliance, whatever the corporate motivation, should have been and continue to be a visible agenda item for employees in almost all company departments~\cite{ico_preparing_2018}. People in Finance and IT would be aware of the cost of additional IT information security expenditure and the potential size of fines. Staff in Human Resources and Customer Service would be aware of personal data handling requirements and subject access requests. Executives in Sales and Marketing would be aware of purpose limitations and the need to gain consent to promotional campaigns. 
We are interested in testing this commitment since public statements of investment and compliance are cheap. We conjecture:

\noindent \textbf{Hypothesis 4}: Companies have responded to GDPR and made changes.

\subsection{Employee awareness of their employer's Data Protection regulator}

It is reasonable to assume the in-house Data Protection Officer (DPO) will know the identity of their DPA since they will be required to communicate with it. However it is less clear if staff, in general, know their DPA and even more so if the DPO function is outsourced to an external provider, which is quite a common business practice for SMEs~\cite{dataguard_data_2022}.

Under article 57.1~\cite{intersoftconsulting_general_2018}, DPAs also have a duty to engage in activities furthering `the awareness of controllers and processors of their obligations` i.e. the individuals and companies that handle personal data. Making companies aware of their responsibilities would seem obvious if they are expected to be held accountable for their actions. However, it is also the nexus of the two schools of enforcement which are (punitive) deterrence or (advisory) compliance. 

Hodges~\citeyear{hodges_delivering_2018} asserts that DPA awareness is essential if it is to play a leadership role where the emphasis is on the expertise, authority and influence of the DPA. Hijmans~\citeyear{hijmans_how_2018} partly agrees but argues that for enforcement of the regulatory framework to be successful, regulators should have sufficient resources and capability to issue strong sanctions. The GDPR includes both compliance and deterrence approaches. %

Data on employee awareness of their employer's GDPR regulator is inconclusive. Deloitte, the management consultancy, Deloitte~\citeyear{deloitte_deloitteukriskgdprsixmonthson_2018} found 57\% of respondents indicated their organisations had received regulatory requests from their DPA. Later in the same year, STAR delivered a report to the EU that found \enquote{Most DPAs neither conduct specific research aimed at establishing levels of SME awareness nor general awareness of the GDPR}~\cite{barnard-wills_report_2019}. %

If companies have invested in training, then employees should be familiar with at least the name of their company's regulator and with the size of the fines their company could face. However, given the uncertainty, we hypothesise:

\noindent \textbf{Hypothesis 5}: Employees lack awareness of the GDPR regulator at work.

\subsection{Employee perception of benefit of the GDPR to their employer}

Literature on the benefits of the GDPR to business is somewhat limited. Studies focus on the technical, financial and human resources struggles that companies encounter in complying with the regulation~\cite{tikkinen-piri_eu_2018b}. Training employees is challenging, including even data scientists. Larsson and Lilja~\citeyear{larsson_gdpr_2019} argue that the GDPR will provide some opportunities for other businesses, like legal experts, lawyers, consultants in digital strategy and professionals in analytics. Poritskiy et al.~\citeyear{poritskiy_benefits_2019a} find the main benefits identified include increased confidence and legal clarification whilst the main challenges include the execution of audits to systems and processes and the application of the right to erasure. %

Almeida Teixeira et al.~\cite{almeidateixeira_critical_2019b}, in their systematic literature review, find the literature reflects on benefits that organisations may achieve by implementing the GDPR including proper data management; use of data analytics; cost reduction; and increase in reputation and competitiveness. This review was conducted in 2019 and looked back at literature that is pre-GDPR mainly in practice.

More recently, Buckley et al.~\citeyear{buckley_it_2022} looked at the actual impact of the GDPR on companies after three years. They found that the GDPR had made companies more privacy-aware, spurred modernisation of their data processes and justified upgrade investments to their security infrastructure. At the same time, they found evidence that it had made new business development harder due to restrictions on requesting and holding personal data, increased direct and indirect compliance costs, left companies confused at times due to grey areas of law and exposed companies to burdensome subject access requests.

Many of these benefits and complications may be invisible or above the pay grade of employees which is why we hypothesise:

\noindent \textbf{Hypothesis 6}: Employees have seen little benefits to their company from GDPR.

\subsection{The research goal is the consumer/employee perception of the GDPR}

Westin's work on different attitudes to privacy~\cite{kumaraguru_privacy_a} and, Acquisti's studies ~\citeyear{acquisti_economics_2016,acquisti_privacy_2005,acquisti_privacy_2015,acquisti_what_2013} on the economics of privacy help us to begin to understand how individuals rationalise the importance and effort they treat protecting their data. People instinctively know there are trade-offs.

With the shift to the internet and the datafication of society, it has created what Shoshanna Zuboff describes as \enquote{a wholly new subspecies of capitalism in which profit drove from the unilateral surveillance and modification of human behaviour}~\citeyear{zuboff_surveillance_2019}. Policymakers have struggled with finding the right balance between competing interests and the GDPR is their latest attempt at squaring the circle.

The purpose of the GDPR is to protect individuals and the data that describes them and to ensure the organisations that collect that data act responsibly. The respondents in our study have lived with the GDPR for four years and have seen it from the inside and outside. They have lived with the hassle of cookie consent notices as consumers and lived with knock-on effects at work. Knowing what they know, we ask the key research question: \enquote{do you think it is worth it?}

\subsection{Summary}
\label{sec:summary}
The literature review has revealed six hypotheses for study:

\begin{enumerate}
    \item Consumers are aware and knowledgeable about the GDPR.
    \item Consumers lack awareness and knowledge about the regulator.
    \item Consumers feel their privacy is better since GDPR was introduced.
    \item Companies have responded to GDPR and made changes.
    \item Employees lack awareness of the GDPR regulator at work.
    \item Employees have seen little benefits to their company from GDPR.
\end{enumerate}

These allow us to study if individuals exposed to GDPR both as employees and as consumers think it is worth it. 

\section{Methods}

The hypotheses underlying the research question lend themselves to qualitative and quantitative analysis. Interview and experimental methods were considered and discounted. Recruiting individuals at scale who have been in continuous employment for over five years from diverse organisations is, unfortunately infeasible. Thus, a survey-based method was both a sensible and a realistic option. 

An early key decision was to limit the survey to the UK. The GDPR may be the same across the EU but the composition of the national regulators and how they implement the regulation are very different for historic reasons. Expanding it to mainland Europe may seem an attractive opportunity to consider a broader cross-cultural perspective but it also came at the cost of introducing too many variables into the study.

\subsection{Design}

The survey was developed in three phases: a test to check interest, a test to check potential population size, and the final study. Participants were recruited via Prolific, an on-demand platform for connecting researchers with volunteers worldwide. The data was collected using the Qualtrics survey platform between the 30th of May and the 16th of June 2022.

In phase \#1, N=10, we used a fast one-minute survey to test the strength of interest in the topic, as we noted some research topics lay ignored for weeks on the platform.
In phase \#2, N=273, we used a longer three-minute survey to confirm there were enough individuals with relevant experiences on Prolific to warrant a full study. %
To ensure respondents had worked pre-and post-GDPR in the workplace, we pre-screened respondents to have at least 5 years of tenure with the same organisation. We also asked if they had heard of the initials GDPR. %
If they answered the initials were unfamiliar to them (only 7\% of respondents), they were paid, thanked and dropped from the survey. It did provide a measure of unfamiliarity or basic unawareness of the GDPR. %
We found respondents answered the survey suspiciously quickly, so we redesigned the survey to add more nonsense and attention checks, repeated and reworded some questions to measure response consistency. The final survey design can be found in Appendix~\ref{app:survey}.

Based on the responses of phase \#2, we expect to find a medium-sized effect ($\approx0.5$) in the main study. For one-tailed repeated measures t-tests at a significance criterion of $\alpha = 0.05$, the minimum sample size is 45 participants. To give room for Bonferroni corrections a sample of 90+ should allow us to achieve statistically significant and generalizable results.

Thus, in phase \#3, we recruited N=102 participants. A representative demographic distribution of the UK was enforced in sampling from the phase \#2 database to make the research findings more generalisable.  The final sample shows an appropriate distribution in terms of gender, age and education compared to census data and consists of 51 female and 51 male respondents with an average age of 45 years for both sexes. Answering the final survey took 10 minutes on average. Seven participants failed exactly 1 of the 7 nonsense, attention and consistency checks. No one failed more than one, so we did not exclude any responses from our analysis. A statistical analysis of mouse and keyboard browser events showed that participants paused and answered questions thoughtfully. This proves we have high-quality responses. %

\subsection{Data Analysis}
The survey consisted of open, closed, slider and multiple-choice questions on a 7-point Likert scale. It focused on the six hypotheses before finishing with the central research question. While the main part of the survey was framed neutrally, we ended with a series of provocative statements to flush out their emotional reaction to the GDPR.

The qualitative analysis of the free text responses was informed by the Braun and Clarke six-step process~\citeyear{braun_using_2006} and the Williams and Moser art of coding and thematic exploration~\citeyear{williams_art_2019}. The first author coded all the data, following an open-axial-selective coding process. An open, primarily inductive process was used to develop an initial codebook. The codes were discussed and refined with the second author in weekly meetings. Any inter-coder differences of interpretation were resolved by discussion. This process eventually led to the themes presented in this work. The codebooks and statistical distribution are in Appendix~\ref{app:responses}.%

The quantitative analysis was executed in Python. The precise statistical tests for each question are described in the Analysis and Results section. The data and reproducible analysis pipeline is available at osf.io\footnote{\url{https://osf.io/rkvz9/?view_only=5f68dc6f7f4a494fad2ef5f8c6eff862}}.

\subsection{Ethical considerations}
\label{sec:ethicalConsiderations}

The authors' departmental Research Ethics Committee approved this study. The online survey is designed to include pseudonymity, confidentiality and informed consent. The study does not identify individual participants. We do not ask questions that could identify the organisations (or the individuals themselves). The participants were aware of the research's purpose, the researchers involved, and their role in it. Participants were offered compensation at a rate of £10 per hour for participating.

\section{Analysis and Results}

\begin{table*}[htb!]
    \caption{Confusion matrix comparing the participant's guesses for the name of UK's GDPR regulator between the pre- and main-study (which were 8 weeks apart)}
    \label{tab:Q16Confusion}
    \centering
    \begin{tabular}{l@{\hspace{\sLPsep}}C{2.4cm}@{\hspace{\sLPsep}}C{2.4cm}@{\hspace{\sLPsep}}C{2.1cm}@{\hspace{\sLPsep}}C{1.8cm}@{\hspace{\sLPsep}}} \toprule
\diagbox[height=0.8cm,width=5cm]{Main Study}{Pre Study} & The British Privacy Authority (BPA) & The Data Protection Authority (DPA) & The Information Commissioners Office (ICO) & The Office of Data (OfDat) \\ \midrule
The Data Protection Agency (DPA) & 0.0 & 20.6 & 2.9 & 1.0 \\
The Data Protection Authority (DPA) & 0.0 & 23.5 & 3.9 & 0.0 \\
The Information Commissioners Office (ICO) & 0.0 & 12.7 & 34.3 & 0.0 \\
The Office of Data (OfDat) & 1.0 & 0.0 & 0.0 & 0.0 \\ \bottomrule
\end{tabular}

\end{table*}

\begin{table*}[htb!]
    \caption{Answers to the question `Which of the following roles is the regulator expected to do?'}
    \label{tab:Q28Answers}
    \centering
    \begin{tabular}{lccc}\toprule
    Roles of the regulator                                        & No     & Unsure & Yes   \\ \midrule
Give advice to members of the public                          & 10.8\% & 31.4\% & 57.8\%\\
Give guidance to companies about their obligations            &  2.0\% &  7.8\% & 90.2\%\\
Maintain a public register of data controllers                &  3.9\% & 45.1\% & 51.0\%\\
Deal with concerns/complaints raised by members of the public &  2.0\% &  8.8\% & 89.2\%\\
Fine companies for proven data misuse                         &  2.9\% & 11.8\% & 85.3\%\\
Fine companies for data breaches                              &  2.0\% & 13.7\% & 84.3\%\\ \bottomrule

    \end{tabular}
\end{table*}

\subsection{Background demographics}

In the final phase \#3 of the study, most participants were from large (250 to 2,499 employees) and very large companies (2,500+). The Operations/Manufacturing and Customer Service departments accounted for just over 50\% of the sample (see Appendix~\ref{app:responses} for details).

\subsection{Hypothesis 1: Consumers are aware and knowledgeable about the GDPR}
\label{sec:hypothesis1}

In the larger phase \#2 survey, 93\% of respondents confirmed awareness of GDPR or the General Data Protection Regulation. Only those who acknowledged familiarity were invited to the subsequent main study.

\begin{figure}[htb!]
    \centering
    \includegraphics[trim={0.2cm 0.3cm 0.2cm 0.2cm},clip,width=\columnwidth]{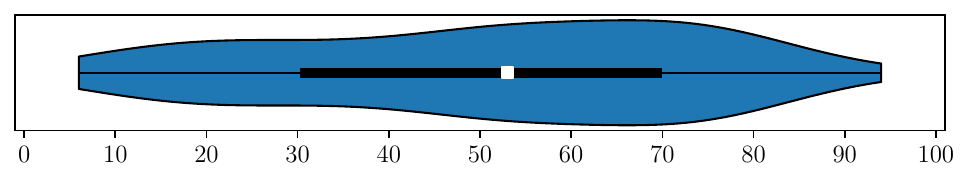}
    \caption{Violin plot of participants self-evaluated knowledge of GDPR consumer rights.}
    \label{fig:q24Violin}
\end{figure}

Regarding the question `How well do you know what rights GDPR gives you as a consumer?', we employed a slider scale from 0 (nothing) to 100 (expert) for more precise quantification. The average score was 50.6, with a median of 53. Notably, the distribution in Figure~\ref{fig:q24Violin} hints at two distinct populations—one less confident in their GDPR knowledge. A Kolmogorov-Smirnov test supports this, rejecting the null hypothesis of a single normal distribution ($p=0$, $s=1.00$).

\begin{figure}[htbp!]
    \centering
    \includegraphics[trim={0.2cm 0.3cm 0.2cm 0.2cm},clip,width=\columnwidth]{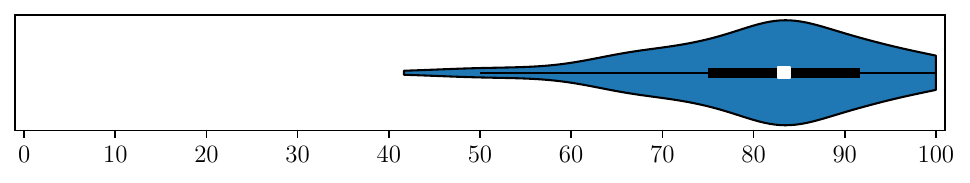}
    \caption{Violin plot showing the percentage of questions correctly answered about consumer rights.}
    \label{fig:q25Violin}
\end{figure}

Respondents were then presented with eight statements, of which four were correct regarding consumer rights, to test their depth of knowledge of the GDPR. The answer options were yes, no or unsure. While the averaged scores per individual are pretty high, only 59\% of respondents got all the positive statements correct, and only 20\% got all the negative statements correct, i.e. correctly identified the incorrect statements (Figure~\ref{fig:q25Violin}).

We conclude there is a high awareness and knowledge of the GDPR. People may lack confidence that they know their consumer rights under the GDPR at a high level but are more sure-footed at a detailed level.  Participants scored high on recognising legitimate rights but were unsure when presented with made-up rights. 

\subsection{Hypothesis 2: Consumers lack awareness and knowledge about the regulator}
\label{sec:hypothesis2}

We assessed consumer awareness of the UK GDPR regulator by repeating questions from the phase \#2 survey. Participants were asked to identify the regulator from five possible answer options. In the survey, 38\% correctly guessed the Information Commissioner’s Office (ICO), and 47\% did so in the subsequent main study. Follow-up analysis revealed that those who initially chose the ICO remained consistent, while those who initially selected the DPA options wavered. Overall, we found no significant learning effect. Refer to Table~\ref{tab:Q16Confusion} for detailed confusion analysis.

This was followed by an open question `What is/are the main purpose(s) of the GDPR regulator?' Table~\ref{tab:Q27Topics} in the appendix shows a topic analysis performed by two researchers using inductive coding to classify the textual responses. The top four purposes were to monitor companies' compliance, protect data (security), issue fines, and ensure against data misuse.

Next, respondents were asked about the regulator’s expected roles from a list of six statements. All of them are true. Table~\ref{tab:Q28Answers} shows the results. 29\% identified all correct statements, and an additional 30\% recognised 5 out of 6.
We conducted repeated-measures adjusted chi-squared tests to assess the significance of these results. The first test rejected the idea that responses (Yes / Unsure / No) were randomly distributed (with $p < 10^{-8}$ for all statements). The second set of tests, focusing on Yes / Unsure responses, also showed no statistical significance (with $p<0.01$) except for ‘Maintain a public register of data controllers’ ($p=0.544$).

Approximately 45\% of participants reported awareness of companies being fined by the regulator, with no statistically significant correlation found between this awareness and the belief that the regulator should fine companies (Mann-Whitney $U=5.5$, $p=0.82$). Interestingly, among those aware of fines, 53\% could not recall the identity of the fined companies. Among those who could recall, the top three mentioned were Facebook, British Airways, and Talk Talk (a UK mobile phone operator).

Given less than half could recognise the ICO from a shortlist, we conclude moderate consumer awareness exists regarding the regulator’s identity, with slightly higher awareness of its role and actions.

\subsection{Hypothesis 3: Consumers feel their privacy is better since GDPR was introduced}
\label{sec:hypothesis3}

In different parts of the survey, participants responded to three Likert scale statements related to GDPR (Table~\ref{tab:H3Answers}). All of the responses were overwhelmingly positive. No statistically significant relationship was found between these answers and participants’ company size or department.

\begin{table*}[htb!]
    \caption{Questions relating to Hypothesis 3.}
    \label{tab:H3Answers}
    \centering
\begin{tabular}{p{\sLQ}@{\hspace{\sLPsep}}C{\sLP}@{\hspace{\sLPsep}}C{\sLP}@{\hspace{\sLPsep}}C{\sLP}@{\hspace{\sLPsep}}C{\sLP}@{\hspace{\sLPsep}}C{\sLP}@{\hspace{\sLPsep}}C{\sLP}@{\hspace{\sLPsep}}C{\sLP}}\toprule%
H3 Questions                                                 & Strongly disagree & Disagree & Mildly disagree & Neither agree or disagree & Mildly agree & Agree  & Strongly agree\\ \midrule
GDPR means makes me feel more in control of personal my data &  2.0\%            &  3.9\%   &  3.9\%          & 14.7\%                    &  9.8\%       & 42.2\% & 23.5\%        \\
GDPR is good for the consumer                                &  1.0\%            &  1.0\%   &  1.0\%          &  8.8\%                    &  4.9\%       & 46.1\% & 37.3\%        \\
GDPR has improved privacy                                    &  1.0\%            &  2.9\%   &  6.9\%          &  9.8\%                    & 11.8\%       & 47.1\% & 20.6\%        \\ \midrule \\[-1.9em]
Aggregate distribution & \vspace{0.2\sLP}\hspace*{0.2cm}\includegraphics[width=8\sLP]{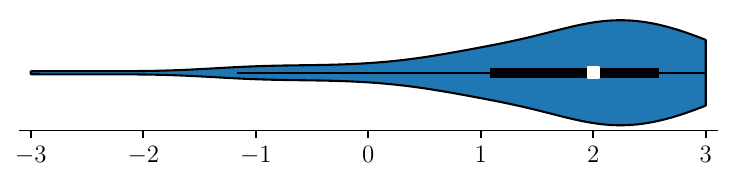}\vspace{-1em} \\ \bottomrule
\end{tabular}
\end{table*}

We calculated a composite score for Hypothesis 3 by averaging individual responses (weighted from -3 to +3 based on their position on the Likert scale). The mean composite score is 1.67 with a standard deviation of 1.2. A t-test ($t=13.94$, $p<10^{-24}$) confirms that this average significantly differs from a mean of zero. Our conclusion: consumers strongly perceive improved privacy since the introduction of GDPR.

\subsection{Hypothesis 4: Companies have responded to GDPR and made changes}
\label{sec:hypothesis4}

\begin{figure}[htb!]
    \centering
    \includegraphics[trim={0.2cm 0.3cm 0.2cm 0.2cm},clip,width=\columnwidth]{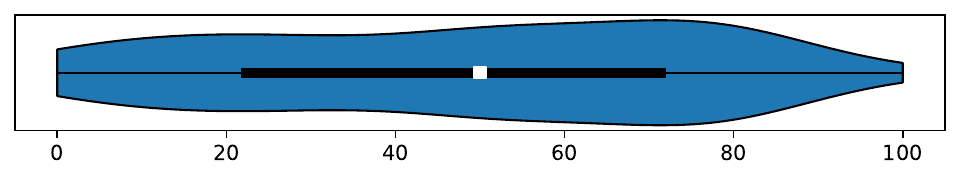}
    \caption{Distribution of answers to `How well do you know what your company has to do in order to comply with GDPR?' on a scale of 0--100.}
    \label{fig:q23Violin}
\end{figure}

When asked to self-evaluate, `How well do you know what your company has to do in order to comply with GDPR?' on a slider scale with 0=`I know nothing' and 100=`I am an expert', the average was 48, the median was 50, and STD was 27 (see Figure~\ref{fig:q23Violin}). %
The distribution appears bimodal, indicating both less knowledgeable and expert individuals. A Kolmogorov-Smirnov test rejects the null hypothesis of a single normal distribution with $p<10^{-100}$ ($s=0.95$).

Respondents were presented with seven statements under the question `Which of the following are rules that a company must comply with when handling personal data under GDPR?' and asked to answer yes, no or unsure. See Table~\ref{tab:Q26Answers} for the results. 

\begin{table}[htb!]
    \caption{Real and made-up rules a company must comply with when handling personal data under GDPR. All obligations bar the 5th are true.}
    \label{tab:Q26Answers}
    \centering
    \begin{tabular}{p{4.7cm}ccc}\toprule
    GDPR obligations                                & No     & Unsure & Yes   \\ \midrule
Fair, lawful and transparent use only                &  1.0\% &  4.9\% & 94.1\%\\
Specific and explicit purpose                        &  6.9\% & 14.7\% & 78.4\%\\
Limited to only what is necessary and relevant       &  5.9\% &  6.9\% & 87.3\%\\
Data must be kept up to date                         &  5.9\% & 14.7\% & 79.4\%\\
Data can be kept for longer than necessary           & 74.5\% & 17.6\% &  7.8\%\\
Data must be kept safe and secure                    &  1.0\% &  0.0\% & 99.0\%\\
Must be made available to national security if asked & 13.7\% & 49.0\% & 37.3\%\\ \bottomrule

    \end{tabular}
\end{table}

We conducted two sets of multiple-comparison adjusted chi-squared tests. The first tested whether responses (Yes/ Unsure/No) could be randomly distributed. This was rejected with p < 0.001 for all statements. The second set focused on Yes/Unsure responses, and again, all are statistically significantly different from random apart from `Must be made available to national security if asked', which has $p=0.2$. Participants score high on knowledge of individual company obligations, with some uncertainty regarding the national security exemption.

Respondents were offered 10 statements on how their employer company had responded to the GDPR. Table~\ref{tab:Q7maindata} in the appendix shows the results. Six of these were also asked in the shorter pilot.%

\begin{table*}[htb!]
    \caption{Questions relating to Hypothesis 5: Employee lack awareness of the GDPR regulator at work.}
    \label{tab:H5Answers}
    \centering
\begin{tabular}{p{\sLQ}@{\hspace{\sLPsep}}C{\sLP}@{\hspace{\sLPsep}}C{\sLP}@{\hspace{\sLPsep}}C{\sLP}@{\hspace{\sLPsep}}C{\sLP}@{\hspace{\sLPsep}}C{\sLP}@{\hspace{\sLPsep}}C{\sLP}@{\hspace{\sLPsep}}C{\sLP}}\toprule%
H5 Questions                                                                                              & Strongly disagree & Disagree & Mildly disagree & Neither agree or disagree & Mildly agree & Agree  & Strongly agree\\ \midrule
The ICO does pop up occasionally in discussions at work                                                    & 23.5\%            & 25.5\%   & 12.7\%          & 13.7\%                    & 13.7\%       &  7.8\% &  2.9\%        \\
In work discussions, the ICO is well regarded                                                              &  4.9\%            & 10.8\%   &  2.9\%          & 61.8\%                    &  6.9\%       &  9.8\% &  2.9\%        \\
Staff have been informed that the company could face big fines by the ICO for data misuse or data breaches & 11.8\%            & 11.8\%   & 12.7\%          & 15.7\%                    &  7.8\%       & 24.5\% & 15.7\%        \\ \midrule \\[-1.9em]
Aggregate distribution & \vspace{0.2\sLP}\hspace*{0.2cm}\includegraphics[width=7.8\sLP]{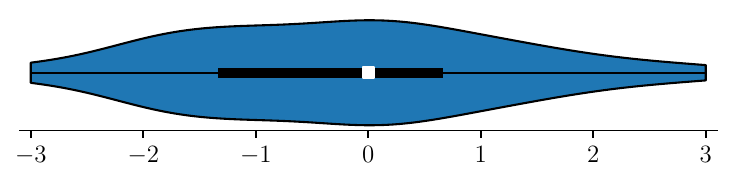}\vspace{-1em} \\ \bottomrule
\end{tabular}
\end{table*}

\begin{figure}[htb!]
    \centering
    \includegraphics[trim={0.2cm 0.3cm 0.2cm 0.2cm},clip,width=\columnwidth]{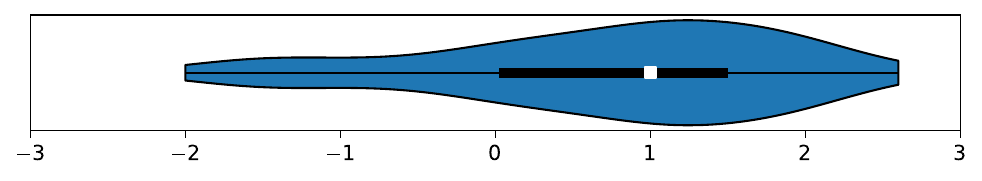}
    \caption{Average observed change in the company due to GDPR.}
    \label{fig:Q7Changes}
\end{figure}

We calculated a composite score for observed organisational changes by weighting each individual's response from -3 through to +3 based on their position on the Likert scale (reversing scales as required) and averaging it over the 10 statements. Figure~\ref{fig:Q7Changes} displays this distribution, with a mean 0.71, std $1.11$. Rejecting the null-hypothesis of a normal distribution with mean of 0 (1~sample t-test with statistic $=6.46$, $p<10^{-8}$), suggests that people have indeed observed changes in behaviour due to the GDPR.

\begin{figure}[htb!]
    \centering
    \includegraphics[trim={0.2cm 0.3cm 0.2cm 0.2cm},clip,width=\columnwidth]{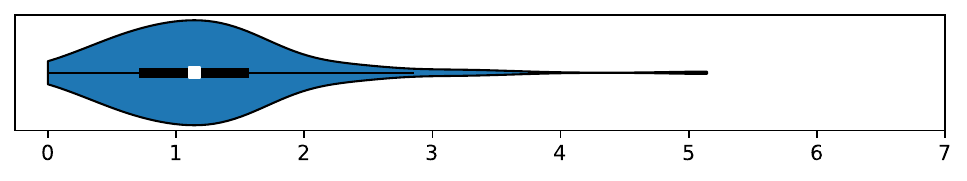}
    \caption{Average absolute difference between Likert responses between the pilot and main study for questions relating to observed changes due to the GDPR.}
    \label{fig:Q7Stability}
\end{figure}

Finally, we compared the scores from the phase 2 pilot and the main study. Figure~\ref{fig:Q7Stability} shows a violin plot of the absolute difference in Likert response scores for questions asked in both studies. Wilcoxon signed-rank tests reveal no significant differences ($p < .01$) in participants' responses across repeated questions in the main survey, conducted 8 weeks later. The non-absolute average, with a mean of 0.09, indicated a minimal change in the time between the pilot and the main study. Overall, people's perceptions of changes in their company have remained remarkably stable.

We conclude our sample believes their employers have responded to the GDPR and observed changes. While they may lack confidence that they know GDPR compliance requirements in theory, their high correct scores on specific questions demonstrate knowledge in practice.

\subsection{Hypothesis 5: Employees lack awareness of the GDPR regulator at work}
\label{sec:hypothesis5}

After ensuring participants knew that the ICO was the UK GDPR regulator, participants were asked to respond to three statements regarding the visibility, reputation and punitive powers of the ICO in their workplace. Table~\ref{tab:H5Answers} shows the questions and results.

The survey shows that the ICO is not a topic of conversation in the office; people have no opinion about its reputation, but they are aware their employer is liable to fines for data misuse or data breaches. We calculated a composite score for Hypothesis 5 by weighting each individual's response from -3 through to +3 depending on where the answer sat on the Likert scale and averaging it over the three questions. The mean is $-0.23$ with a standard deviation of $1.41$. We cannot reject the null hypothesis that this distribution is drawn from a Normal distribution with mean 0 (one sample t-test with statistic$=-1.61$, $p=0.11$). It is possible that participants were answering randomly to this question. We concluded that employee awareness of the GDPR regulator in the office is mixed at best. 

\subsection{Hypothesis 6: Employees have seen little benefits to their company from GDPR.}
\label{sec:hypothesis6}

We analysed this question in three stages: First, we asked participants to what extent their job had changed due to the GDPR, followed by open questions about the advantages and disadvantages of the GDPR. Then, on the next page, participants were asked to what extent they agreed with eight impact statements about the GDPR, and finally, we asked them to judge if the GDPR was good for their company. All answers can be found in Table~\ref{tab:H6Answers}, and we will discuss each of the previous parts in turn next.

\begin{table*}[htb!]
    \centering
    \caption{Questions relating to Hypothesis 6. The second and all subsequent statements are about the impact of GDPR on the respondents company, Questions 2--5 are about negative aspects of GDPR, while questions 6--9 are about positive aspects.}
    \label{tab:H6Answers}
\begin{tabular}{p{\sLQ}@{\hspace{\sLPsep}}C{\sLP}@{\hspace{\sLPsep}}C{\sLP}@{\hspace{\sLPsep}}C{\sLP}@{\hspace{\sLPsep}}C{\sLP}@{\hspace{\sLPsep}}C{\sLP}@{\hspace{\sLPsep}}C{\sLP}@{\hspace{\sLPsep}}C{\sLP}}\toprule%
H6 Questions                                                            & Strongly disagree & Disagree & Mildly disagree & Neither agree or disagree & Mildly agree & Agree  & Strongly agree\\ \midrule
The requirements of GDPR have made my job harder and/or more cumbersome &  7.8\%            & 22.5\%   & 14.7\%          & 21.6\%                    & 25.5\%       &  3.9\% &  3.9\%        \\ \midrule
Less bureaucracy                                                        & 11.8\%            & 23.5\%   & 20.6\%          & 36.3\%                    &  4.9\%       &  2.0\% &  1.0\%        \\
More compliance costs                                                   &  1.0\%            &  2.9\%   & 12.7\%          & 33.3\%                    & 27.5\%       & 17.6\% &  4.9\%        \\
More cyber-security costs                                               &  2.9\%            &  4.9\%   &  4.9\%          & 28.4\%                    & 27.5\%       & 25.5\% &  5.9\%        \\
Receive more freedom of information requests                            &  7.8\%            & 11.8\%   &  5.9\%          & 42.2\%                    &  9.8\%       & 13.7\% &  8.8\%        \\ \midrule
Better data security                                                    &  2.9\%            &  1.0\%   &  2.0\%          & 11.8\%                    & 16.7\%       & 52.9\% & 12.7\%        \\
Better awareness of the importance of data privacy practices            &  2.0\%            &  1.0\%   &  0.0\%          & 16.7\%                    & 19.6\%       & 41.2\% & 19.6\%        \\
Less up-to-date customer databases                                      &  5.9\%            & 29.4\%   & 20.6\%          & 34.3\%                    &  3.9\%       &  3.9\% &  2.0\%        \\
Better trust and confidence in the company brand                        &  2.9\%            &  2.0\%   &  3.9\%          & 38.2\%                    & 11.8\%       & 33.3\% &  7.8\%        \\ \midrule
GDPR is good for my company                                             &  0.0\%            &  0.0\%   &  5.9\%          & 28.4\%                    &  6.9\%       & 45.1\% & 13.7\%        \\ \bottomrule
\end{tabular}
\end{table*}

\begin{table*}[htb!]
\centering
    \caption{Questions relating to the main research question: \enquote{do you think it is worth it?}}
    \label{tab:H7Answers}
\begin{tabular}{p{\sLQ}@{\hspace{\sLPsep}}C{\sLP}@{\hspace{\sLPsep}}C{\sLP}@{\hspace{\sLPsep}}C{\sLP}@{\hspace{\sLPsep}}C{\sLP}@{\hspace{\sLPsep}}C{\sLP}@{\hspace{\sLPsep}}C{\sLP}@{\hspace{\sLPsep}}C{\sLP}}\toprule%
H7 Questions                                                 & Strongly disagree & Disagree & Mildly disagree & Neither agree or disagree & Mildly agree & Agree  & Strongly agree\\ \midrule
GDPR means makes me feel more in control of personal my data &  2.0\%            &  3.9\%   &  3.9\%          & 14.7\%                    &  9.8\%       & 42.2\% & 23.5\%        \\
GDPR is good for the consumer                                &  1.0\%            &  1.0\%   &  1.0\%          &  8.8\%                    &  4.9\%       & 46.1\% & 37.3\%        \\
GDPR is good for my company                                  &  0.0\%            &  0.0\%   &  5.9\%          & 28.4\%                    &  6.9\%       & 45.1\% & 13.7\%        \\
On balance, GDPR is worth it                                 &  1.0\%            &  1.0\%   &  4.9\%          & 13.7\%                    &  5.9\%       & 50.0\% & 23.5\%        \\ \bottomrule
\end{tabular}
\end{table*}

Initially, participants were fairly evenly split about the effect of the GDPR on their job. This continued to be the case when asked about the advantages and disadvantages of the GDPR, where on average, participants agree with mean 0.58 (std $=0.92$) that there are negatives to the GDPR, and with mean 1.16 (std $=0.98$) that there are positives to the GDPR. Both distributions are statistically significantly different from a normal of mean 0 (with statistics of $6.3$ and $12.0$, and $p=10^{-9}$ and $p=10^{-21}$ respectively). Still, the positives are statistically considerably stronger than the negatives (related samples t-test, statistic $=-4.47$, $p=2.1\times10^{-5}$).

This even split contrasts strongly with the final statement, with almost no participant disagreeing that \emph{GDPR is good for their company}. The ordering may have been a potential biasing factor and/or engaging in the exercise may have helped people to form a more concrete opinion (see also Section~\ref{sec:discussion-unformed-opinion}).

Through free-text responses, we explored the pros and cons further. First, respondents were asked to identify the biggest disadvantage for their company. Responses (full codebook in the appendix in Table~\ref{tab:Q35Topics}) were categorized into five clusters: no observed disadvantage, increased bureaucratic processes, higher costs, constraints on customer data, and miscellaneous complaints. The most common response was no observed changes or disadvantages, with some attributing this to existing robust processes. The top themes included increased bureaucracy and paperwork, as well as more time-consuming processes. Respondents also mentioned ongoing compliance costs, staff training requirements, and constraints on data collection for marketing purposes as significant drawbacks. Other cited disadvantages included responding to freedom of information requests, internal information sharing difficulties, and uncertainty regarding inadvertent GDPR breaches.

Respondents were asked to identify the biggest advantage of the GDPR for their company. The responses (full codebook in the appendix in Table~\ref{tab:Q36Topics}) can be categorized into three clusters: better data protection and security, clearer rules, and third, no discernible advantage to their company. Under data protection, respondents linked improved information security and enhanced trust in their company. This applied to client data and their own personal data held by their employer. Some cited transparency and compliance as enhancing their company's brand. Clearer rules led to standardized processes, improved employee training, and better handling of personal data, potentially protecting the company from fines. Some respondents also noted benefits such as the GDPR incentivizing data upkeep, discarding out-of-date information and reducing storage costs. 

We conclude people recognise the benefits to their companies but are not blind to the disbenefits of the GDPR.

\subsection{Research question: GDPR: Is it worth it?}

Participants were asked to respond to four statements about GDPR, including the central research question. The overwhelmingly positive responses can be seen in Table~\ref{tab:H7Answers}: it appears that when forced to recall and judge the positives and negatives of GPDR, they conclude that it is good not just for them, but also for their employer.

\subsection{A regression model based on the dual professional-consumer perspective}

\begin{figure}[htb!]
    \begin{adjustbox}{width=\columnwidth,center}
    \tikzstyle{tikzfig}=[baseline=-0.25em,scale=0.5]
\pgfdeclarelayer{edgelayer}
\pgfdeclarelayer{nodelayer}
\pgfsetlayers{background,edgelayer,nodelayer,main}

\tikzstyle{arrow}=[->, thick]
\tikzstyle{square}=[fill=white, draw=black, shape=rectangle, align=center]

\tikzstyle{none}=[inner sep=0mm]

\begin{tikzpicture}
	\begin{pgfonlayer}{nodelayer}
		\node [style=square] (0) at (0, 0) {Consumer feels privacy is better}; %
		\node [style=square] (1) at (-3.7, 1.75) {Knowledge of GDPR}; %
		\node [style=square] (2) at (-2.75, 3) {Knowledge of Roles of Regulator};  %
		\node [style=square] (3) at (2.5, 3) {Positive impacts};
		\node [style=square] (4) at (3.5, 1.75) {Negative impacts};
		\node [style=square] (5) at (-4, -1.25) {Not scared \&\\ do bare minimum}; %
		\node [style=square] (6) at (-2.8, -2.8) {Not scared because fines unlikely}; %
		\node [style=square] (7) at (0, -4) {GDPR is worth it}; %
		\node [style=square] (8) at (2.8, -2.5) {More hassle than it is worth it}; %
		\node [style=square] (9) at (3.5, -1.25) {Good for my company}; %
	\end{pgfonlayer}
	\begin{pgfonlayer}{edgelayer}
		\draw [style=arrow] (2) to node[above=1.5mm,right=0mm] {$0.232^{***}$} (0) ;
		\draw [style=arrow] (1) to node[left=1mm] {$0.220^{***}$} (0);
		\draw [style=arrow] (3) to node[above=5mm,left=-5mm] {$0.551^{***}$} (0);
		\draw [style=arrow] (4) to node[right=2mm] {$-0.205^{***}$} (0);
		\draw [style=arrow] (0) to node[right=3mm] {$0.664^{***}$} (9);
		\draw [style=arrow] (0) to node[below=5mm,left=-4.8mm] {$-0.371^{***}$} (8);
		\draw [style=arrow] (0) to node[below=13mm,right=0.6mm] {$0.829^{***}$} (7);
		\draw [style=arrow] (0) to node[below=4mm,right=-4.5mm] {$-0.391^{***}$} (6);
		\draw [style=arrow] (0) to node[left=1mm] {$-0.269^{***}$} (5);
	\end{pgfonlayer}
\end{tikzpicture}
    \end{adjustbox}\vspace{0.2em}
    \caption{Model of our findings, based on 6 regression models (one inputs model, five output models). All coefficients are statistically significant at $p<0.001$.}
    \label{fig:H3Graph}
\end{figure}
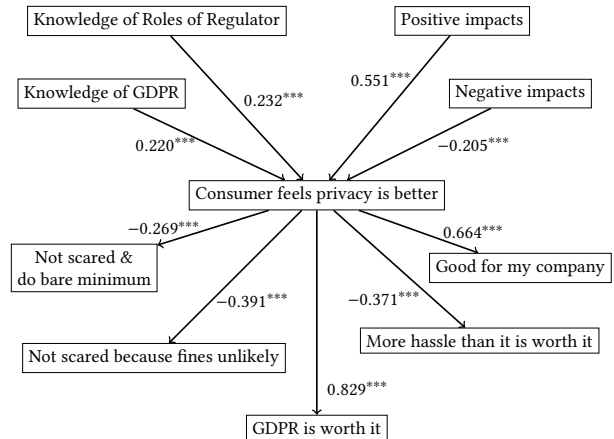

Given the unique dual perspective of our participants, we explored potential dependencies between our hypotheses. Using 20-fold cross-validated step-wise linear regression models, we identified the smallest set of questions (or composite scores that represent our hypotheses) that maximize model explainability. Our analysis, conducted in Python using scipy and the statsmodels package, can be found in the online supplementary materials. The full regression tables are available in Appendix~\ref{app:regression}. Based on this analysis, we propose a new model for understanding the perceptions and influences of the GDPR (Figure~\ref{fig:H3Graph}).

We found that consumers’ perception of improved privacy (measured through Hypothesis 3) is pivotal for our outcome variables. Several moderating factors influence this view: knowledge of the GDPR (Hypothesis 1), understanding regulator roles (Hypothesis 3), and observing positive impacts on their company due to the GDPR (Hypothesis 6).
Our main research question, ‘Is GDPR worth it?’, is well-explained by a model based solely on ‘Consumer feels privacy is better,’ achieving an impressive $R^2=0.687$. Our participants recognize the impact of GDPR on their privacy and value it accordingly.

\section{Discussion}

This research examined the informed individual's perception of the GDPR. This is important because we can gauge buy-in and learn what works when considering new privacy regulations.
Several high-level themes can be drawn from the results:

\subsection{High consumer awareness and knowledge of the GDPR}

The H1 results reveal a strong awareness of GDPR among participants, with 93\% acknowledging it in the phase \#2 survey, a marked improvement over previous EU surveys. Our research tallies with the literature that people learn about GDPR from news, employer training, and cookie consent notices. Departments like HR, IT, Marketing, and Legal exhibit higher awareness than others, possibly due to its greater impact on their work. 

While participants aren’t spontaneously confident about their GDPR rights, they recognize consumer rights when prompted. Notably, they understand the right to be informed and the right to request data copies. They are less confident recognising fabricated rights---registering high unsure scores---but that is probably the right reaction.
Equally, while participants aren't spontaneously confident about their employer's GDPR compliance obligations, they scored high overall, with the exception of the national security exemption, which is hardly common knowledge.

\subsection{Respondents lacked a formed opinion}
\label{sec:discussion-unformed-opinion}

The sequencing of the questions was designed to avoid bias before posing the central research question: ‘Is GDPR worth it?’ Initially, respondents were hazy about GDPR and its impact on their job. However, once prompted with specific questions about GDPR, the regulator, and observed repercussions at work, they finished the survey with a positive evaluation of GDPR. 
We speculate participants may default to imprecise gut feelings unless prompted to consider its specific benefits and drawbacks. Future data protection surveys may improve response quality by giving participants the space to develop an opinion.

\subsection{GDPR has driven changes}
The results from H4 prove people have seen changes at work. This shows the GDPR is working.
We can be confident the answers are `solid' because they are very similar to the answers they gave to the same questions two months beforehand in the phase 2 pilot.
In particular, they have observed how personal data is handled more carefully, and they have received regular training on the risk of fines due to data misuse or data breaches.
More generally, they agreed more than disagreed with all the prompted observed changes. 
The change that received the highest uncertain score was `My company collects less personal data than before', but even then, more people agreed than disagreed.

People recognise the upsides (improved data security) and downsides (bureaucracy, time, cost) of these GDPR-driven changes for their employers and for them personally.
This latter point regarding cybersecurity tallies with earlier research. However, we bring an original contribution behind how our participants evaluate this.
For their employer, people think better information security means less chance of fines. We speculate they may translate this as better job security for themselves.
For the employee, better information security makes them feel their own data is handled more carefully by their own employer. We speculate they may project this expectation onto other companies.

\subsection{Perceptions of privacy have improved}
The results for H3 show people feel their privacy is better since the introduction of the GDPR in 2018. This is an important and positive finding since comparative empirical surveys are sparse. The ICO surveys annual changes in trust and confidence scores rather than privacy per se. Other perception research looks at control~\cite{presthus_threeyear_2021a}, choice and risk perceptions~\cite{bornschein_effect_2020}.

\subsection{The profile of the regulator may not matter}
The results for H2 \& H5 show people are not very familiar with the ICO as consumers or as employees. It registers modest name recognition. That said, ICO awareness has improved since the EU Eurobarometer survey in 2019. Back then, only 21\% of the UK knew its identity, whereas 38\% recognised the ICO in phase \#2.
People believe its role is twofold---to monitor compliance, including data security and data misuse, and to issue fines.
This matches some of the ICO mission statements. People see the ICO as more company-facing and less consumer-facing.
This may not matter since the ICO achieves its objectives via controls through companies.
If the people that matter, i.e. DPOs and senior management, rather than ordinary employees are aware of the ICO, then it may not hinder its effectiveness.

\subsection{Regulator = Enforcer}
People's expectation of enforcement is complex. Unlike the enforcer to advisor spectrum outlined in the literature review, the general population have firm expectations of their regulator as an aggressive compliance-led defender of their rights rather than an advice-led consultant to them and their employers. They see the regulator as there to punish and fine non-compliant businesses.
However, half of them cannot remember any company being fined. Of the half that could, half could not remember the names of the culprits---so how important are fines really to their perception of the regulator and the regulation?
Our regression analysis suggests that people are most likely to believe companies are scared of GDPR fines if they feel privacy has improved because they have observed changes in their own employer and are aware of the compliance obligations of companies.

\subsection{GDPR is worth it if...}

Our regression analysis suggests the following mental models at play:
People believe the GDPR is worth it because they feel their privacy is better since the GDPR was introduced.
People are most likely to say this if they are confident they know their consumer rights, know the regulator's powers and have observed first-hand the mix of positive and negative changes at work.

People are most likely to think the GDPR is good for their company and not too much hassle if they have seen more positive and fewer negative changes to it and, curiously, are not too knowledgeable of the role of the regulator and the compliance obligations. A sort of goldilocks situation~\cite{_goldilocks_}: they see more positive than negative changes and don't regard the regulator as too powerful or demanding.

\subsection{Implications}

We examine the implications of these themes across theoretical, managerial, and policymaker/regulator levels.

Theoretically, we question the sustainability of GDPR-inspired changes observed in this study. Will there be compliance decay over time, considering the competition for business focus from newer regulations? The GDPR benefited from massive publicity at launch, but that was five years ago. There are grounds for hope. Recent experiences suggest that EU data regulation often reinforces compliance in other areas, potentially mitigating decay. Additionally, new data protection regulations overseas, inspired by the GDPR, may reinvigorate its relevance.

At a managerial level, our research suggests that constant awareness and knowledge training of the GDPR can lead to unforeseen effects. Raised expectations among employees for high data hygiene practices from their employers may drive companies to promote their GDPR credentials in order to reassure staff and customers and foster trust in their brand.

Our research also prompts consideration of the optimum positioning for a regulator. If the ICO's focus is primarily on corporate compliance rather than consumer protection, this has implications for policymakers. For instance, should the UK government direct the ICO to issue more fines to reinforce their deterrent effect on corporates?

The positive perception of the GDPR among those who comply and implement it suggests valuable lessons for future policymaking. Incorporating early feedback and buy-in from a dual professional-consumer sample population may enhance the development of new regulations in this field.

\subsection{Limitations and future work}

Although the UK GDPR is virtually identical to the EU GDPR, the findings from a UK-only sample may not be applicable to other EU countries, especially regarding regulator-specific results due to differences in national regulator competencies and resources.
As we wanted to ensure non-identifiable responses, we cannot estimate the diversity of companies studied, and there may have been multiple responses from single companies. 
By expanding the sample size, future work could investigate if participants' views are influenced or different based on their industry sector or country regulator. 
Just before submission, the ICO published a survey with a larger sample size than ours, corroborating our findings with regard to comparable questions. Another research path would be to explore how new complementary data-related regulations reinforce each other and influence consumer perceptions.

\section{Conclusion}

In the UK, GDPR awareness is high, and consumers understand their rights. They perceive improved data protection and personal data control since the introduction of the GDPR. While the regulator’s identity awareness is lower, participants recognize its role in upholding rights and imposing fines. This may become a point of dissatisfaction in the long term, as participants struggled to recall companies that had actually been fined.

Interestingly, employees view GDPR as good for their companies because it protects customer data and their personal data. Despite recognizing the overheads (people, process, and technical), they believe GDPR clarifies compliance requirements on their employer and what it has to do to avoid being fined. They appreciate that their employers (and, by extension, other companies) must be more conscientious in handling and securing personal data. In summary, while GDPR may be viewed as an imposition, participants still think it is worth it. These insights have important implications for policymakers and regulators who may wish to emulate this public support for future regulation roll-outs.

\section*{Funding and Disclosure statement}
\ifanonymous Redacted for anonymous submission. \else 
Gerard Buckley is supported by UK EPSRC grant no. EP/S022503/1. Ingolf Becker is supported by UK EPSRC grant no. EP/W032368/1. The authors report there are no competing interests to declare.
\fi

\balance
\bibliographystyle{ACM-Reference-Format}
\bibliography{P21.bib}

\clearpage
\nobalance
\onecolumn
\appendix

\section{Tables of survey responses}
\label{app:responses}
\begin{table}[htb!]
    \caption{Organisation size compared with division}
    \centering
    \label{tab:size-vs-division}
    \begin{tabular}{l@{\hspace{0.4em}}C{1.2cm}@{\hspace{0.4em}}c@{\hspace{0.4em}}C{1.1cm}@{\hspace{0.4em}}C{1.2cm}@{\hspace{0.4em}}C{1.1cm}@{\hspace{0.4em}}C{1.2cm}@{\hspace{0.4em}}C{1.6cm}@{\hspace{0.4em}}C{1.5cm}@{\hspace{0.4em}}C{1.3cm}@{\hspace{0.4em}}|c@{\hspace{0.4em}}}\toprule
     & Customer Service & Finance & General management & Human resources & Informa\-tion Technology & Legal/ Risk/ Compliance & Marketing/ Product Manager/ Digital Marketer & Operations/ Manufacturing & Sales/ Business Development & Sum \\ \midrule
Micro 1--9 & 0 & 1 & 1 & 0 & 0 & 0 & 0 & 2 & 2 & 6 \\
Small 10--49 & 3 & 1 & 2 & 0 & 2 & 0 & 0 & 1 & 2 & 11 \\
Medium 50--249 & 2 & 0 & 3 & 0 & 2 & 1 & 3 & 7 & 1 & 19 \\
Large 250--2,499 & 5 & 1 & 3 & 1 & 3 & 1 & 0 & 5 & 1 & 20 \\
Very Large 2,500+ & 15 & 3 & 6 & 0 & 7 & 1 & 0 & 11 & 3 & 46 \\ \midrule
Sum & 25 & 6 & 15 & 1 & 14 & 3 & 3 & 26 & 9 & 102 \\ \bottomrule

    \end{tabular}
\end{table}

\begin{table}[htb!]
\parbox{.45\linewidth}{
    \caption{Perceived main purpose of the GDPR regulator.}
    \centering
    \label{tab:Q27Topics}
    \begin{tabular}{lcc}\toprule
    Topics                               & Count & \%      \\ \midrule
Monitor compliance                         & 59    & $57.8\%$\\
Protect data                               & 32    & $31.4\%$\\
Fine                                       & 11    & $10.8\%$\\
Misuse                                     & 11    & $10.8\%$\\
Don't know                                 & 8     & $ 7.8\%$\\
Investigate breaches                       & 8     & $ 7.8\%$\\
Support consumers                          & 8     & $ 7.8\%$\\
Support companies                          & 7     & $ 6.9\%$\\
Take action                                & 7     & $ 6.9\%$\\
Uphold consumer rights                     & 6     & $ 5.9\%$\\
Protect people                             & 5     & $ 4.9\%$\\
Ensure accuracy                            & 3     & $ 2.9\%$\\
Audit                                      & 2     & $ 2.0\%$\\
Deal with complaints                       & 2     & $ 2.0\%$\\
Enable individual to control personal data & 2     & $ 2.0\%$\\
Impartial                                  & 2     & $ 2.0\%$\\
Transparency                               & 2     & $ 2.0\%$\\
Can make SAR                               & 1     & $ 1.0\%$\\
EU wide standard                           & 1     & $ 1.0\%$\\
It is a con                                & 1     & $ 1.0\%$\\
Promote data privacy                       & 1     & $ 1.0\%$\\ \bottomrule

    \end{tabular}
}\quad\parbox{.45\linewidth}{
    \caption{Codes and frequency of advantages of the GDPR to the respondents' company.}
    \label{tab:Q36Topics}
    \centering
    \begin{tabular}{lcc}\toprule
    Topics                         & Count & \%      \\ \midrule
Don't know                     & 25    & $24.5\%$\\
Better care handling data      & 12    & $11.8\%$\\
Better personal data security  & 12    & $11.8\%$\\
Better data security           & 11    & $10.8\%$\\
Clear guidance                 & 7     & $ 6.9\%$\\
Trust signal to employee       & 7     & $ 6.9\%$\\
Employee peace of mind         & 6     & $ 5.9\%$\\
Better data management         & 5     & $ 4.9\%$\\
Enforces compliance            & 5     & $ 4.9\%$\\
Trust signal to consumer       & 5     & $ 4.9\%$\\
Collect less data              & 4     & $ 3.9\%$\\
More transparency              & 4     & $ 3.9\%$\\
More awareness of penalties    & 3     & $ 2.9\%$\\
No advantage to company        & 3     & $ 2.9\%$\\
Training                       & 3     & $ 2.9\%$\\
Employee privacy awareness     & 2     & $ 2.0\%$\\
More training                  & 2     & $ 2.0\%$\\
More trust                     & 2     & $ 2.0\%$\\
Reason to refuse               & 2     & $ 2.0\%$\\
Accountability                 & 1     & $ 1.0\%$\\
Employee control over own data & 1     & $ 1.0\%$\\
GDPR support industry          & 1     & $ 1.0\%$\\
More up to date                & 1     & $ 1.0\%$\\
Protects company               & 1     & $ 1.0\%$\\
Protects consumer              & 1     & $ 1.0\%$\\
Reassurance                    & 1     & $ 1.0\%$\\
Tailored advertising           & 1     & $ 1.0\%$\\
Uniform training               & 1     & $ 1.0\%$\\ \bottomrule

    \end{tabular}}
\end{table}

\begin{table}[htb!]
    \caption{Perceived response by the participant's employer to the GDPR.}
    \label{tab:Q7maindata}
    \centering
\begin{tabular}{p{\sLQ}@{\hspace{\sLPsep}}C{\sLP}@{\hspace{\sLPsep}}C{\sLP}@{\hspace{\sLPsep}}C{\sLP}@{\hspace{\sLPsep}}C{\sLP}@{\hspace{\sLPsep}}C{\sLP}@{\hspace{\sLPsep}}C{\sLP}@{\hspace{\sLPsep}}C{\sLP}}\toprule%
Main Questions                                                                                                   & Strongly disagree & Disagree & Mildly disagree & Neither agree or disagree & Mildly agree & Agree  & Strongly agree\\ \midrule
GDPR has changed how I do my job at work                                                                            &  4.9\%            &  9.8\%   &  7.8\%          & 14.7\%                    & 26.5\%       & 28.4\% &  7.8\%        \\
I have not been put on GDPR-related privacy and cybersecurity training courses at work                              & 23.5\%            & 20.6\%   &  9.8\%          &  3.9\%                    &  6.9\%       & 17.6\% & 17.6\%        \\
I've noticed new and/or widened roles and responsibilities at work that were inspired by GDPR                       &  6.9\%            & 11.8\%   &  6.9\%          & 16.7\%                    & 28.4\%       & 24.5\% &  4.9\%        \\
I've noticed no new privacy and security software at work since GDPR                                                & 13.7\%            & 27.5\%   & 19.6\%          & 11.8\%                    &  6.9\%       & 14.7\% &  5.9\%        \\
I've noticed my employer has not changed their behaviour when handling personal customer  data since GDPR           & 22.5\%            & 32.4\%   & 16.7\%          & 13.7\%                    &  5.9\%       &  4.9\% &  3.9\%        \\
My employer has not communicated to staff that it could face big fines for data misuse or data breaches under GDPR. & 30.4\%            & 26.5\%   & 13.7\%          &  7.8\%                    &  2.0\%       & 13.7\% &  5.9\%        \\
My company is more transparent about how it uses customer data than before                                          &  3.9\%            &  5.9\%   &  4.9\%          & 23.5\%                    & 30.4\%       & 23.5\% &  7.8\%        \\
My company takes more care when handling personal data than before                                                  &  2.0\%            &  2.9\%   &  2.9\%          & 20.6\%                    & 18.6\%       & 37.3\% & 15.7\%        \\
My company collects less personal data than before                                                                  &  4.9\%            & 13.7\%   &  9.8\%          & 44.1\%                    &  9.8\%       & 15.7\% &  2.0\%        \\
My company has not made material changes since GDPR                                                                 & 14.7\%            & 31.4\%   & 16.7\%          & 19.6\%                    &  7.8\%       &  3.9\% &  5.9\%        \\ \bottomrule
\end{tabular}
\end{table}

\begin{table}[htb!]
    \caption{Codes and frequency of disadvantages of the GDPR to the respondents company.}
    \label{tab:Q35Topics}
    \centering
    \begin{tabular}{lcc}\toprule
    Topics                             & Count & \%      \\ \midrule
Bureaucracy                        & 23    & $22.5\%$\\
More processes                     & 18    & $17.6\%$\\
Time-consuming                     & 16    & $15.7\%$\\
No disadvantage                    & 14    & $13.7\%$\\
Don't know                         & 11    & $10.8\%$\\
No change                          & 7     & $ 6.9\%$\\
Risk uncertain                     & 7     & $ 6.9\%$\\
Extra costs                        & 6     & $ 5.9\%$\\
Hinders marketing                  & 6     & $ 5.9\%$\\
Limitation of use                  & 6     & $ 5.9\%$\\
None                               & 6     & $ 5.9\%$\\
Business development drawback      & 5     & $ 4.9\%$\\
More security                      & 5     & $ 4.9\%$\\
Hinders customer service           & 4     & $ 3.9\%$\\
More workload                      & 4     & $ 3.9\%$\\
Training                           & 4     & $ 3.9\%$\\
Abuse of rights                    & 3     & $ 2.9\%$\\
Answering SARs                     & 2     & $ 2.0\%$\\
Staff resistence                   & 2     & $ 2.0\%$\\
More careful handling              & 1     & $ 1.0\%$\\
Prefer to ignore GDPR and pay fine & 1     & $ 1.0\%$\\ \bottomrule

    \end{tabular}
\end{table}

\clearpage
\section{Regression analysis}
\label{app:regression}

\setlength{\lmQ}{4.4cm}

\setlength{\lmP}{1.4cm}

\setlength{\lmPsep}{0.2cm}

\begin{table*}[htb!]
    \caption{Linear Models for outcome variables associated with the organisation. The rows are the independent variables, which were selected stepwise in a cross-validated manner to minimize model error.}
    \label{tab:lm-company}
    \centering
    \begin{tabular}{p{\lmQ}@{\hspace{\lmPsep}}C{\lmP}@{\hspace{\lmPsep}}C{\lmP}@{\hspace{\lmPsep}}C{\lmP}@{\hspace{\lmPsep}}C{\lmP}@{\hspace{\lmPsep}}C{\lmP}@{\hspace{\lmPsep}}C{\lmP}@{\hspace{\lmPsep}}C{\lmP}@{\hspace{\lmPsep}}C{\lmP}@{\hspace{\lmPsep}}} \toprule
    \diagbox[height=1cm,width=0.9\lmQ]{Independent}{Dependent}                 & Not scared \& do bare minimum & Scared \& changed behaviour & Not scared because fines unlikely & More hassle than worth & Good for my company & Knowledge of compliance & Knowledge of obligations & Observed changes \\ \midrule
\multirow{2}{=}{Consumer feels privacy is better}         & -0.269*** &           & -0.391*** & -0.371*** & 0.664***  &          &          &            \\ %
                 & (0.093)   &           & (0.090)   & (0.106)   & (0.074)   &          &          &            \\
\multirow{2}{=}{Awareness of ICO at work} & -0.211**  &           &           &           &           &          &          & 0.190**    \\ %
                 & (0.091)   &           &           &           &           &          &          & (0.081)    \\
\multirow{2}{=}{Knowledge of compliance}          &           &           &           &           &           &          & 0.273*** & 0.221***   \\ %
                 &           &           &           &           &           &          & (0.090)  & (0.083)    \\
\multirow{2}{=}{Knowledge of GDPR}           &           &           &           &           &           & 0.648*** &          &            \\ %
                 &           &           &           &           &           & (0.063)  &          &            \\
\multirow{2}{=}{Knowledge of rights}         &           &           &           &           & -0.186**  &          & 0.324*** &            \\ %
                 &           &           &           &           & (0.074)   &          & (0.090)  &            \\
\multirow{2}{=}{Knowledge of obligations}         &           &           &           &           &           &          &          & 0.154**    \\ %
                 &           &           &           &           &           &          &          & (0.071)    \\
\multirow{3}{=}{Knowledge of Roles of Regulator}         &           &           &           & 0.173**   &           &          &          &            \\ %
                 &           &           &           & (0.080)   &           &          &          &            \\ \\
\multirow{2}{=}{Negative impacts}  &           &           &           & 0.231***  &           &          &          & 0.254***   \\ %
                 &           &           &           & (0.077)   &           &          &          & (0.069)    \\
\multirow{2}{=}{Positive impacts}  &           & 0.399***  &           & -0.338*** &           &          &          & 0.310***   \\ %
                 &           & (0.091)   &           & (0.100)   &           &          &          & (0.077)    \\
\multirow{3}{=}{Changes in company behaviour}        &           &           &           &           &           & 0.298*** &          &            \\ %
                 &           &           &           &           &           & (0.063)  &          &            \\ \\
\multirow{2}{=}{Participant certainty}      & 0.253***  &           & 0.197**   &           &           &          &          & -0.170**   \\ %
                 & (0.089)   &           & (0.090)   &           &           &          &          & (0.069)    \\ \midrule
R-squared        & 0.268     & 0.159     & 0.227     & 0.451     & 0.451     & 0.662    & 0.195    & 0.570      \\
R-squared Adj.   & 0.246     & 0.151     & 0.212     & 0.429     & 0.440     & 0.655    & 0.179    & 0.543      \\ \bottomrule

    \end{tabular}\vspace{-0.5em}
\end{table*}

\begin{table*}[htb!]
    \caption{Linear Model for outcome variables associated with the individual.}
    \label{tab:lm-consumer}
    \centering
    \begin{tabular}{p{5cm}@{\hspace{\lmIPsep}}C{\lmIP}@{\hspace{\lmIPsep}}C{\lmIP}@{\hspace{\lmIPsep}}C{\lmIP}@{\hspace{\lmIPsep}}} \toprule
    
\diagbox[width=3.5cm]{Independent}{Dependent}    & More hassle than worth it & Consumer feels privacy is better & On balance, GDPR is worth it \\ \midrule
\multirow{2}{=}{Consumer feels privacy is better}         & -0.602*** &           & 0.829***   \\
                 & (0.074)   &           & (0.056)    \\
\multirow{2}{=}{Knowledge of GDPR}           &           & 0.220***  &            \\
                 &           & (0.072)   &            \\
\multirow{2}{=}{Knowledge of Roles of Regulator}         &           & 0.232***  &            \\
                 &           & (0.069)   &            \\
\multirow{2}{=}{Negative impacts}  & 0.233***  & -0.205*** &            \\
                 & (0.074)   & (0.069)   &            \\
\multirow{2}{=}{Positive impacts}  &           & 0.551***  &            \\
                 &           & (0.072)   &            \\ \midrule
R-squared        & 0.459     & 0.547     & 0.687      \\
R-squared Adj.   & 0.448     & 0.528     & 0.684      \\ \bottomrule

    \end{tabular}\vspace{-5em}
\end{table*}

\clearpage

\ifanonymous 
\includepdf[pages=1-2,scale=0.8,nup=2x1,frame,pagecommand={\section{Survey}\label{app:survey}},offset=0cm 0]{surveypdf-cropped-annonymised-flattened.pdf}
\includepdf[pages=3-,scale=0.8,nup=2x1,frame,pagecommand={},offset=0cm 0]{surveypdf-cropped-annonymised-flattened.pdf}
\else 
\includepdf[pages=1-2,scale=0.8,nup=2x1,frame,pagecommand={\section{Survey}\label{app:survey}},offset=0cm 0]{surveypdf-cropped.pdf}
\includepdf[pages=3-,scale=0.8,nup=2x1,frame,pagecommand={},offset=0cm 0]{surveypdf-cropped.pdf}
\fi

\end{document}